\documentclass[aps,prl,twocolumn,superscriptaddress,floatfix,notitlepage]{revtex4-1}

\usepackage{amsthm}

\usepackage{graphicx,psfrag,color}
\usepackage{mathbbol,amsmath,amsfonts,amssymb,bm,dsfont,braket}
\usepackage{wasysym}
\usepackage[section]{placeins}
\usepackage{multirow}

\def\i{{\sf i}}

\usepackage{verbatim}

\usepackage{array}
\newcolumntype{x}[1]{>{\centering\hspace{0pt}}p{#1}}
\usepackage[normalem]{ulem}

\begin{document}

\title{Zeeman Pumping of Higgs Bosons in the Balian--Werthamer State}

\begin{abstract}
Based on equations of motion of an SO(5) pseudo-spin, we demonstrate a quantum quench protocol using the magnetic pulse to excite an $\textit{undamped}$ heavy Higgs boson in the Balian--Werthamer superfluid (or superconductor). To achieve that, it is essential to include the dipolar interaction in the effective Hamiltonian and to calculate the ground state self-consistently. The pumped heavy Higgs mode has the twisted angular momentum $J=2$ with the projection $J^{\,}_z=0$ and couples to a well-known light Higgs mode ($J=1$, $J^{\,}_z=0$). The numerical method of 26-point Lebedev quadrature is implemented concretely so as to observe the real-time coupled oscillation.
\end{abstract}

\author{Qiao-Ru Xu}
\affiliation{Institute for Theoretical Sciences, Westlake University, Hangzhou 310030, China}
\affiliation{Institute of Natural Sciences, Westlake Institute for Advanced Study, Hangzhou 310024, China}

\date{\today}
\maketitle

{\it Introduction.---}Nuclear magnetic resonance (NMR) experiments have played a crucial role in identifying the A and B phases of liquid $^3$He as spinful $p$-wave superfluid states \cite{Leggett}, with the former being the Anderson--Brinkman--Morel state and the latter being the Balian--Werthamer (BW) state. For the BW state \cite{BW}, collective modes that usually couple to the NMR signal are spin-wave modes, i.e., Goldstone bosons associated with the spontaneously broken relative spin-orbit rotational symmetry \cite{LeggettSBSOS} of the overall SO(3)$^{\,}_S\times$SO(3)$^{\,}_L\times$U(1)$^{\,}_N$ group. Nonetheless, the joint spin-orbit rotational symmetry \cite{Sauls} is still conserved and the twisted angular momentum $J$ (and its projection $J^{\,}_z$) can still be used to label collective modes of the BW state \cite{J,Hughes}. In the presence of a dipolar interaction, the longitudinal spin-wave mode ($J=1$, $J^{\,}_z=0$) acquires a small mass and becomes a pseudo-Goldstone boson \cite{LeggettEOM}, also known as the light Higgs \cite{Volovik2014,Volovik2016}, which can be described by the Leggett equations of motion (EOM) \cite{LeggettEOM}. However, the Leggett EOM are derived under an adiabatic approximation that filters out high-energy degrees of freedom and thus excludes the existence of heavy Higgs bosons by construction.

\begin{figure}[t]
	\smallskip
	\includegraphics[width=\columnwidth]{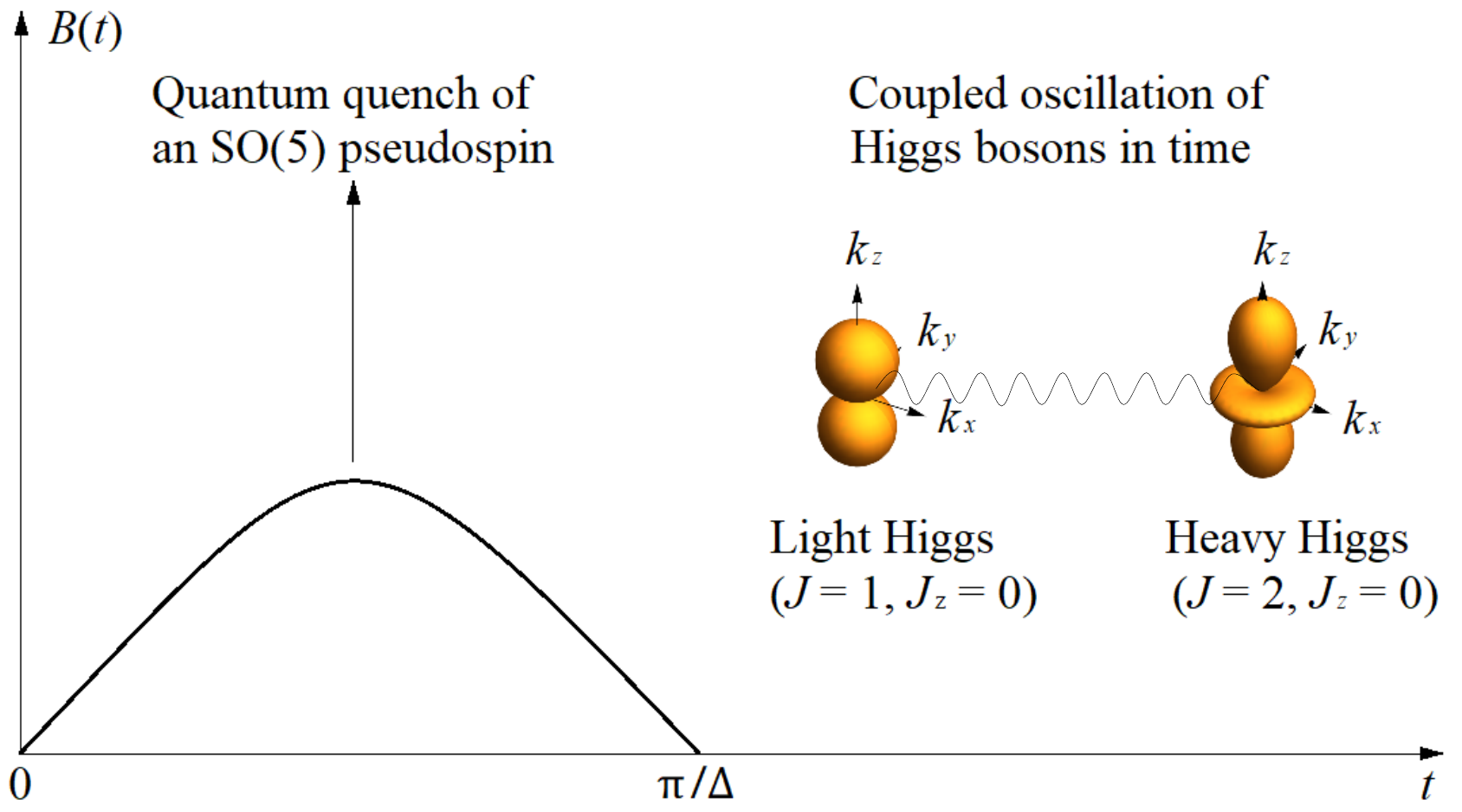}
	\caption{Schematic plot of a quench protocol using the magnetic pulse $\mathbf{B}(t)=B(t)\hat{z}$, with $B(t)\propto \sin(\Delta t)\Theta(t)\Theta(\pi/\Delta-t)$ and $\Theta(t)$ the unit step function. Along with the well-known light Higgs boson ($J=1$, $J^{\,}_z=0$), the coupled heavy Higgs boson ($J=2$, $J^{\,}_z=0$) is also excited.}\label{Coupling}
\end{figure}

Indeed, including spin-wave modes, there are at least 18 collective modes in the BW state, reflected as fluctuations of a $3\times3$ complex order parameter matrix, among which there are totally 14 heavy Higgs modes \cite{Wolfle,Volovik2014,Sauls}. In the long wavelength limit, ten of them are below the continuous spectrum's edge $2\Delta$ and are $\textit{undamped}$ after perturbative quantum quenches \cite{Xu}, in contrast with the oscillation with a power-law decay of the Higgs boson in $s$-wave superconductors \cite{Volkov,Yuzbashyan,Matsunaga,Tsuji,Shimano}. More specifically, half of the ten $\textit{undamped}$ heavy Higgs modes have the energy $\sqrt{8/5}\Delta$ while the other half $\sqrt{12/5}\Delta$ \cite{Wolfle}. The $\sqrt{8/5}\Delta$ Higgs modes and also the spin-wave modes are associated with the real part of the superfluid order parameter fluctuations and can be uniformly described by the EOM of an SO(5) pseudo-spin \cite{Xu,Hasegawa}. In the absence of the dipolar interaction, it has been shown that heavy Higgs modes are not excited in NMR-type quench protocols \cite{Xu} and therefore how to generate a coupling between heavy Higgs modes and spin-wave modes would be essential for practical NMR-type experiments. In this paper, within the SO(5) pseudo-spin formalism, we revisit the interplay between the $\sqrt{8/5}\Delta$ Higgs modes and the spin-wave modes by including the dipolar interaction in the system. It is explicitly demonstrated that a suitable magnetic pulse can induce not only the light Higgs mode but also the $\sqrt{8/5}\Delta$ Higgs mode with $J=2$ and $J^{\,}_z=0$, and they are coupled (see Fig.\,\ref{Coupling}). To kick off, let us start from an effective Hamiltonian written in second quantization.


{\it Effective SO(5) model.---}Consider an effective Hamiltonian of a spinful $p$-wave superfluid in three spatial dimensions \cite{LeggettEOM} as follows 
\begin{align}
\widehat{\mathcal{H}}&=\sum_{\mathbf{k}\alpha}\xi^{\,}_\mathbf{k}c^\dagger_{\mathbf{k}\alpha} c^{\,}_{\mathbf{k}\alpha}+\frac{1}{2}\sum_{\mathbf{k}\mathbf{k'}\alpha\beta}V^{\,}_{\mathbf{k}\mathbf{k'}}c^\dagger_{\mathbf{k}\alpha}c^\dagger_{-\mathbf{k}\beta}c^{\,}_{-\mathbf{k'}\beta}c^{\,}_{\mathbf{k'}\alpha}\nonumber\\
&+\frac{1}{2}\sum_{\mathbf{k}\mathbf{k'}\alpha\alpha'\beta\beta'}V^{\,}_{\alpha\beta\alpha'\beta'}(\mathbf{k},\mathbf{k'})c^\dagger_{\mathbf{k}\alpha}c^\dagger_{-\mathbf{k}\beta}c^{\,}_{-\mathbf{k'}\beta'}c^{\,}_{\mathbf{k'}\alpha'}\nonumber\\
&+\sum_{\mathbf{k}\alpha\alpha'}(-\frac{\gamma\hbar}{2}\mathbf{B})\cdot\boldsymbol{\sigma}^{\,}_{\alpha\alpha'}c^\dagger_{\mathbf{k}\alpha}c^{\,}_{\mathbf{k}\alpha'},
\end{align}
where $\xi^{\,}_\mathbf{k}=\frac{\hbar^2k^2}{2m}-\epsilon^{\,}_F$ is the kinetic energy of spin 1/2 fermions of mass $m$ and momentum $\mathbf{k}$ (with $k=|\mathbf{k}|$) measured from the Fermi energy $\epsilon^{\,}_F$, $c^\dagger_{\mathbf{k}\alpha}$ ($c^{\,}_{\mathbf{k}\alpha}$) is the fermion creation (annihilation) operator with spin $\alpha \in \{\uparrow,\downarrow\}$, $V^{\,}_{\mathbf{k}\mathbf{k'}}=-3g\mathbf{n}\cdot\mathbf{n'}$ is the pairing interaction restricted to the region near the Fermi surface with $g>0$ and $\mathbf{n}=\mathbf{k}/|\mathbf{k}|$, $V^{\,}_{\alpha\beta\alpha'\beta'}(\mathbf{k},\mathbf{k'})=\frac{g^{\,}_D}{2}[3(\hat{\mathbf{q}}\cdot\boldsymbol{\sigma}^{\,}_{\alpha\alpha'})(\hat{\mathbf{q}}\cdot\boldsymbol{\sigma}^{\,}_{\beta\beta'})-\boldsymbol{\sigma}^{\,}_{\alpha\alpha'}\cdot\boldsymbol{\sigma}^{\,}_{\beta\beta'}]$ is the dipolar interaction assumed to be perturbative with $0<g^{\,}_D\ll g$, $\hat{\mathbf{q}}=\frac{\mathbf{k}-\mathbf{k'}}{|\mathbf{k}-\mathbf{k'}|}\approx\frac{\mathbf{n}-\mathbf{n'}}{|\mathbf{n}-\mathbf{n'}|}$, and $\boldsymbol{\sigma}=(\sigma^1,\sigma^2,\sigma^3)$ the Pauli matrices, and finally the last Zeeman term describes magnetic moments of strength $\gamma\hbar/2$ ($\gamma$ the gyromagnetic ratio) in a magnetic field $\mathbf{B}$.

Following the notation of Ref.\,\cite{Xu}, now we introduce the ten generators of the SO(5) algebra. Besides $\mathbf{S}^{\,}_\mathbf{k}=\hat{\mathbf{s}}^{\,}_\mathbf{k}+\hat{\mathbf{s}}^{\,}_\mathbf{-k}$ with $\hat{\mathbf{s}}^{\,}_\mathbf{k}=\frac{1}{2}\sum_{\alpha\alpha'}\boldsymbol{\sigma}^{\,}_{\alpha\alpha'}c^\dagger_{\mathbf{k}\alpha}c^{\,}_{\mathbf{k}\alpha'}$ the spin 1/2 operator and $N^-_\mathbf{k}=\frac{1}{2}(\hat{n}^{\,}_\mathbf{k}+\hat{n}^{\,}_\mathbf{-k})-1$ with $\hat{n}^{\,}_\mathbf{k}=\sum_{\alpha}c^\dagger_{\mathbf{k}\alpha} c^{\,}_{\mathbf{k}\alpha}$ the number operator, the other six generators are $\mathbf{U}^{\,}_\mathbf{k}=\frac{1}{2}(\mathbf{T}^{\,}_\mathbf{k}+\mathbf{T}^\dagger_\mathbf{k})$ and $\mathbf{V}^{\,}_\mathbf{k}=\frac{\i}{2}(\mathbf{T}^{\,}_\mathbf{k}-\mathbf{T}^\dagger_\mathbf{k})$ with $\mathbf{T}^\dagger_\mathbf{k}=\sum_{\alpha\beta}(\i\boldsymbol{\sigma}\sigma^2)^{\,}_{\alpha\beta}c^\dagger_{\mathbf{k}\alpha}c^\dagger_{\mathbf{-k}\beta}$ associated with the creation of spin-triplet Cooper pairs. Then the effective Hamiltonian can be conveniently rewritten in terms of these SO(5) generators 
\begin{align}\label{effective}
	\widehat{\mathcal{H}}
	&=\sum_{\mathbf{k}}\xi^{\,}_\mathbf{k}(N^-_\mathbf{k}+1)
	+\frac{1}{4}\sum_{\mathbf{k}\mathbf{k'}}V^{\,}_{\mathbf{k}\mathbf{k'}}(\mathbf{U}^{\,}_\mathbf{k}\cdot\mathbf{U}^{\,}_\mathbf{k'}+\mathbf{V}^{\,}_\mathbf{k}\cdot\mathbf{V}^{\,}_\mathbf{k'})\nonumber\\
	&+\frac{g^{\,}_D}{4}\sum_{\mathbf{k}\mathbf{k'}}\left[-3(\hat{\mathbf{q}}\cdot\mathbf{U}^{\,}_\mathbf{k})(\hat{\mathbf{q}}\cdot\mathbf{U}^{\,}_\mathbf{k'})-3(\hat{\mathbf{q}}\cdot\mathbf{V}^{\,}_\mathbf{k})(\hat{\mathbf{q}}\cdot\mathbf{V}^{\,}_\mathbf{k'})\right]\nonumber\\
	&+\sum_\mathbf{k}(-\frac{\gamma\hbar}{2}\mathbf{B})\cdot\mathbf{S}^{\,}_\mathbf{k},
\end{align}
where we have used symmetry properties $\mathbf{U}^{\,}_\mathbf{k}=-\mathbf{U}^{\,}_\mathbf{-k}$ and $\mathbf{V}^{\,}_\mathbf{k}=-\mathbf{V}^{\,}_\mathbf{-k}$ during the derivation. Next we adopt the mean-field method to tackle the problem.

{\it Mean-field approximation.---}Within the SO(5) pseudo-spin formalism \cite{Xu}, the pseudo-spin is defined as  $\mathbf{L}^{\,}_\mathbf{k}=[\mathbf{S}^{\,}_\mathbf{k}\, \mathbf{U}^{\,}_\mathbf{k}\, \mathbf{V}^{\,}_\mathbf{k}\, N^-_\mathbf{k}]^\text{T}$ and the associated pseudo-magnetic field is defined as $\mathbf{H}^{\,}_\mathbf{k}=[\mathbf{H}^S_\mathbf{k}\, \mathbf{H}^U_\mathbf{k}\, \mathbf{H}^V_\mathbf{k}\, H^N_\mathbf{k}]$, with  $\mathbf{H}^S_\mathbf{k}=-\gamma\hbar\mathbf{B}$, $H^N_\mathbf{k}=2\xi^{\,}_\mathbf{k}$, and 
\begin{align}
&\mathbf{H}^U_\mathbf{k}=\sum_\mathbf{k'}V^{\,}_{\mathbf{k}\mathbf{k'}}\braket{\mathbf{U}^{\,}_\mathbf{k'}}+\mathbf{R}^U_\mathbf{k},\label{gapeq}\\
&\mathbf{R}^U_\mathbf{k}=g^{\,}_D\sum_\mathbf{k'}\left[-3\left(\hat{\mathbf{q}}\cdot\braket{\mathbf{U}^{\,}_\mathbf{k'}}\right)\hat{\mathbf{q}}\right],\label{Rk}\\
&\mathbf{H}^V_\mathbf{k}=\sum_\mathbf{k'}V^{\,}_{\mathbf{k}\mathbf{k'}}\braket{\mathbf{V}^{\,}_\mathbf{k'}}+\mathbf{R}^V_\mathbf{k},\\
&\mathbf{R}^V_\mathbf{k}=g^{\,}_D\sum_\mathbf{k'}\left[-3\left(\hat{\mathbf{q}}\cdot\braket{\mathbf{V}^{\,}_\mathbf{k'}}\right)\hat{\mathbf{q}}\right].
\end{align}
Then we can write the mean-field Hamiltonian of Eq.\,(\ref{effective}) in a compact form
\begin{align}\label{HMF}
\widehat{H}=\sideset{}{'}\sum_\mathbf{k}(\mathbf{H}^{\,}_\mathbf{k}\cdot\mathbf{L}^{\,}_\mathbf{k}+C)+\frac{1}{2}\mathbf{H}^{\,}_\mathbf{0}\cdot\mathbf{L}^{\,}_\mathbf{0}+\xi^{\,}_\mathbf{0},
\end{align}
where the primed summation is taken over the domain $\{\mathbf{k}|k^{\,}_1>0\}\cup\{\mathbf{k}|k^{\,}_2>0,k^{\,}_1=0\}\cup\{\mathbf{k}|k^{\,}_3>0,k^{\,}_2=0,k^{\,}_1=0\}$ and the constant $C=2\xi^{\,}_\mathbf{k}-\frac{1}{2}\left(\mathbf{H}^U_\mathbf{k}\cdot\braket{\mathbf{U}^{\,}_\mathbf{k}}+\mathbf{H}^V_\mathbf{k}\cdot\braket{\mathbf{V}^{\,}_\mathbf{k}}\right)$.

Let us temporarily turn off both the dipolar interaction and the magnetic field by setting $g^{\,}_D=|\mathbf{B}|=0$. Then $\mathbf{H}^U_\mathbf{k}$ and $\mathbf{H}^V_\mathbf{k}$ reduce to those of Ref.\,\cite{Xu} with $\mathbf{H}^U_\mathbf{k}=2\text{Re\,}\mathbf{d}^*_\mathbf{k}$ and $\mathbf{H}^V_\mathbf{k}=2\text{Im\,}\mathbf{d}^*_\mathbf{k}$, where $\mathbf{d}^{\,}_\mathbf{k}$ is the usual order-parameter vector in the spin space. For the BW state \cite{BW}, we have $\mathbf{d}^{\,}_\mathbf{k}=\Delta R\mathbf{n}$ with $R$ an arbitrary rotation matrix. Without loss of generality, we choose the condition $\Delta \in \mathbb{R}$ so that we have the pseudo-magnetic field $\mathbf{H}^{\,}_{\mathbf{k}}=[\mathbf{0},\, 2\Delta R\mathbf{n},\, \mathbf{0},\, 2\xi^{\,}_\mathbf{k}]$, being antiparallel with $\braket{\mathbf{L}^{\,}_\mathbf{k}}=[\mathbf{0},\, -(\Delta/E^{\,}_\mathbf{k})R\mathbf{n},\, \mathbf{0},\, -\xi^{\,}_\mathbf{k}/E^{\,}_\mathbf{k}]^\text{T}$ in the ground state, and the single-particle excitation energy $E^{\,}_{\mathbf{k}}=\sqrt{\xi^2_\mathbf{k}+\Delta^2}$. However, we know from Ref.\,\cite{Xu} that no heavy Higgs modes on top of the BW ground state can be excited through quantum quenches by coupling the system to the magnetic field. 

To demonstrate a practical quench protocol, we are now at a right position to turn on the dipolar interaction. As pointed out by Leggett \cite{LeggettEOM} that the true BW state is achieved when $R$ is a rotation through the angle $\theta^{\,}_L=\arccos(-1/4)$ at which the dipolar energy is minimized. However, the problem is that $\mathbf{H}^U_\mathbf{k}=2\Delta R\mathbf{n}$ and $\braket{\mathbf{U}^{\,}_\mathbf{k}}=-(\Delta/E^{\,}_\mathbf{k})R\mathbf{n}$ at $\theta^{\,}_L$ is not a self-consistent solution to Eq.\,(\ref{gapeq}). In order to obtain a convincing evidence of heavy Higgs modes, we need to calculate the ground state self-consistently at least to the first order in $g^{\,}_D/g$, which we will do next.

{\it Self-consistent calculation.---}Let us start from Eq.\,(\ref{gapeq}) and denote $\braket{\mathbf{U}^{\,}_\mathbf{k}}$ as follows
\begin{align}\label{Uk}
\braket{\mathbf{U}^{\,}_\mathbf{k}}=\braket{\mathbf{U}^{(0)}_\mathbf{k}}+(g^{\,}_D/g)\braket{\mathbf{U}^{(1)}_\mathbf{k}}+\mathcal{O}(g^2_D/g^2),
\end{align}
with the zeroth order term $\braket{\mathbf{U}^{(0)}_\mathbf{k}}=-(\Delta/E^{\,}_\mathbf{k})R\mathbf{n}$ and the dipole-induced first order term $(g^{\,}_D/g)\braket{\mathbf{U}^{(1)}_\mathbf{k}}$ to be determined. Then from Eqs.\,(\ref{gapeq}) and (\ref{Rk}) we have
\begin{align}
&\hspace*{-0.2cm}\mathbf{H}^U_\mathbf{k}=2\Delta R\mathbf{n}+\mathbf{D}^{\,}_\mathbf{k}+\mathcal{O}(g^2_D/g^2),\label{HUk}\\
&\hspace*{-0.2cm}\mathbf{D}^{\,}_\mathbf{k}=g^{\,}_D\sum_\mathbf{k'}[-3(\hat{\mathbf{q}}\cdot\braket{\mathbf{U}^{(0)}_\mathbf{k'}})\hat{\mathbf{q}}]+\frac{g^{\,}_D}{g}\sum_\mathbf{k'}V^{\,}_{\mathbf{k}\mathbf{k'}}\braket{\mathbf{U}^{(1)}_\mathbf{k'}}.\hspace*{-0.2cm}\label{Dk}
\end{align}
Meanwhile we know that $\braket{\mathbf{U}^{\,}_\mathbf{k}}=-\frac{1}{2}\mathbf{H}^U_\mathbf{k}/\sqrt{\xi^2_\mathbf{k}+|\frac{1}{2}\mathbf{H}^U_\mathbf{k}|^2}$ and it is straightforward to do a Taylor expansion to the first order in $\mathbf{D}^{\,}_\mathbf{k}$, resulting in
\begin{align}\label{Uk1}
&(g^{\,}_D/g)\braket{\mathbf{U}^{(1)}_\mathbf{k}}
=\frac{\Delta^2(\mathbf{D}^{\,}_\mathbf{k}\cdot R\mathbf{n})}{2E^3_\mathbf{k}}R\mathbf{n}-\frac{1}{2E^{\,}_{\mathbf{k}}}\mathbf{D}^{\,}_\mathbf{k}.
\end{align}

Armed with above notations, we first determine the rotation angle $\theta$ of $R$ that minimizes the total energy, not just the dipolar energy as Leggett did. From Eq.\,(\ref{effective}), up to the first order in $g^{\,}_D$, the total energy can be expressed as $\braket{\widehat{\mathcal{H}}}=E^{\,}_0+(\varepsilon^{\,}_D+\varepsilon^{\,}_K+\varepsilon^{\,}_P)+\mathcal{O}(g^2_D)$, with the BW ground state energy $E^{\,}_0$, the first order dipolar energy $\varepsilon^{\,}_D$, and two other dipole-induced first order terms
\begin{align}
&\varepsilon^{\,}_K=\sum_{\mathbf{k}}\xi^{\,}_\mathbf{k}\braket{N^{-,(1)}_\mathbf{k}},\\
&\varepsilon^{\,}_P=\frac{g^{\,}_D}{2g}\sum_{\mathbf{k}\mathbf{k'}}V^{\,}_{\mathbf{k}\mathbf{k'}}\braket{\mathbf{U}^{(0)}_\mathbf{k}}\cdot\braket{\mathbf{U}^{(1)}_\mathbf{k'}}.
\end{align}
Plugging in $\braket{N^{-,(1)}_\mathbf{k}}=\xi^{\,}_\mathbf{k}\Delta(\mathbf{D}^{\,}_\mathbf{k}\cdot R\mathbf{n})/(2E^3_\mathbf{k})$, the first order Taylor expansion of $\braket{N^-_\mathbf{k}}=-\xi^{\,}_\mathbf{k}/\sqrt{\xi^2_\mathbf{k}+|\frac{1}{2}\mathbf{H}^U_\mathbf{k}|^2}$, it is not hard to prove the equality $\varepsilon^{\,}_P=-\varepsilon^{\,}_K$, which means that the problem of minimizing $\braket{\widehat{\mathcal{H}}}$ reduces to that of Leggett of minimizing $\varepsilon^{\,}_D$, leading to $\theta=\theta^{\,}_L$.

\begin{figure}[t]
	\smallskip
	\includegraphics[width=\columnwidth]{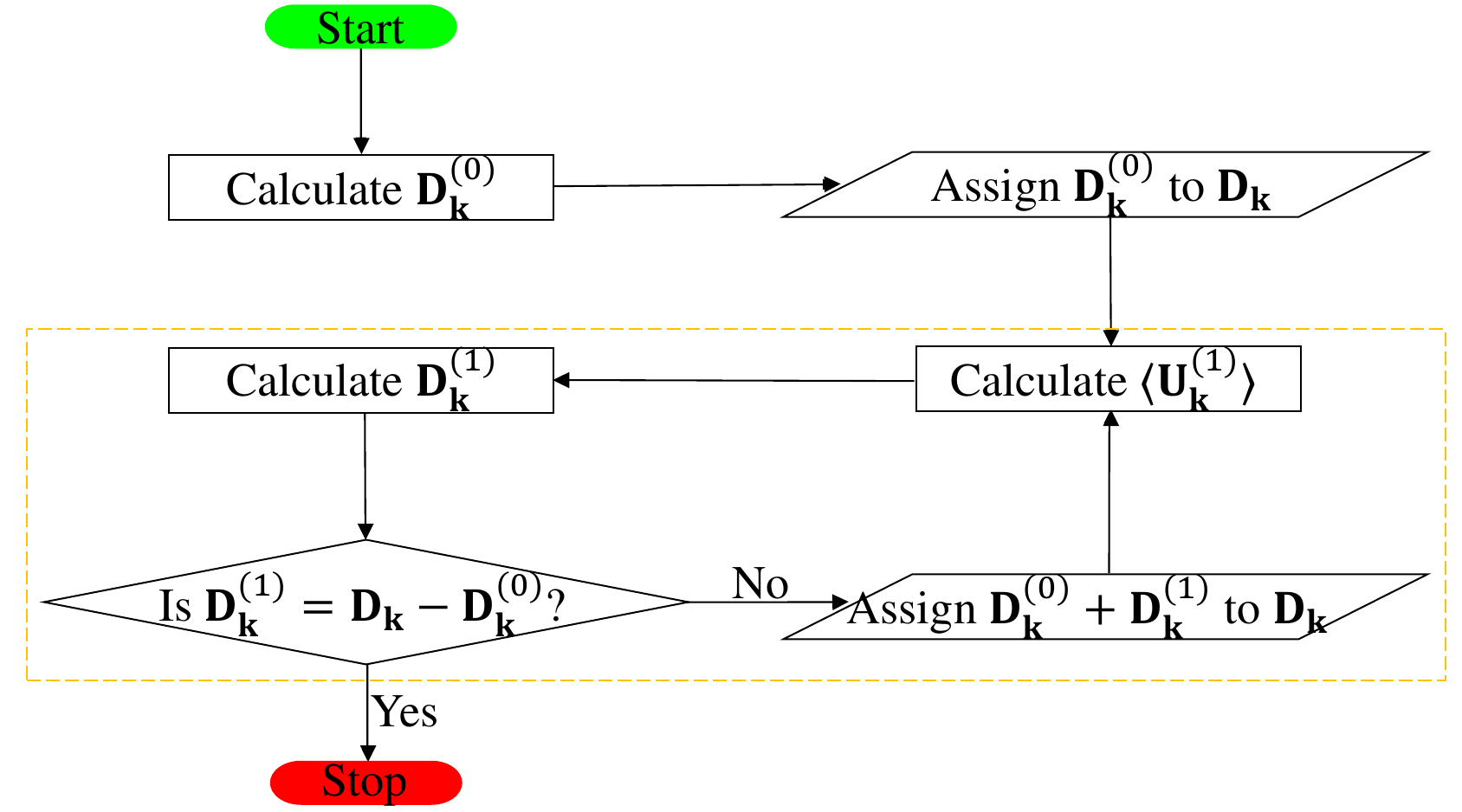}
	\caption{Flowchart of computing $\mathbf{D}^{\,}_\mathbf{k}$ self-consistently. Steps inside the dashed rectangle represent the iteration process of calculating $\braket{\mathbf{U}^{(1)}_\mathbf{k}}$ until $\mathbf{D}^{(1)}_\mathbf{k}$ reaches a convergence. As explained in the main text, $\mathbf{D}^{(0)}_\mathbf{k}$ and $\mathbf{D}^{(1)}_\mathbf{k}$ are the first and second terms of Eq.\,(\ref{Dk}), respectively. When evaluating $\braket{\mathbf{U}^{(1)}_\mathbf{k}}$, Eq.\,(\ref{Uk1}) is used.}\label{Iteration}
\end{figure}

Now we solve for $\mathbf{D}^{\,}_\mathbf{k}$ self-consistently, using the mathematical method of iteration. We see from Eqs.\,(\ref{HUk}) and (\ref{Dk}) that, after switching on the dipolar interaction, $\mathbf{H}^U_\mathbf{k}$ acquires an instantaneous first order correction due to nonzero $\braket{\mathbf{U}^{(0)}_\mathbf{k}}$ in the first term of $\mathbf{D}^{\,}_\mathbf{k}$, which we denote as $\mathbf{D}^{(0)}_\mathbf{k}$. This in turn induces a nonzero $\braket{\mathbf{U}^{(1)}_\mathbf{k}}$ through Eq.\,(\ref{Uk1}) and therefore $\mathbf{H}^U_\mathbf{k}$ acquires another correction due to the second term of $\mathbf{D}^{\,}_\mathbf{k}$, which we denote as $\mathbf{D}^{(1)}_{\mathbf{k}}$ in general or, more specifically, as $\mathbf{D}^{(1)}_{\mathbf{k},j}$ in the $j$'th iteration. This again feeds back to $\braket{\mathbf{U}^{(1)}_\mathbf{k}}$ and the process continues until $\braket{\mathbf{U}^{(1)}_\mathbf{k}}$ and therefore $\mathbf{D}^{\,}_\mathbf{k}$ reach a convergence. Thus, a good starting point of solving Eq.\,(\ref{Dk}) is $\mathbf{D}^{(0)}_\mathbf{k}$ (see Fig.\,\ref{Iteration}). After some calculation, we find that
\begin{align}\label{Dk0}
\hspace*{-0.05cm}\mathbf{D}^{(0)}_\mathbf{k}\hspace*{-0.05cm}=\hspace*{-0.05cm}\frac{g^{\,}_D}{g}2\Delta\hspace*{-0.05cm}\left[\hspace*{-0.05cm}\frac{(\mathbf{n}\cdot R\mathbf{n})\mathbf{n}+R\mathbf{n}}{4}\hspace*{-0.05cm}-\hspace*{-0.05cm}\frac{(\text{Tr}R)\mathbf{n}+R^{-1}\mathbf{n}}{2}\hspace*{-0.05cm}\right]\hspace*{-0.05cm},\hspace*{-0.05cm}
\end{align}
where we have used the substitution $\sum_{\mathbf{k}}=N(0)\int^{\epsilon^{\,}_c}_{-\epsilon^{\,}_c}d\xi^{\,}_\mathbf{k}\int\frac{d\Omega^{\,}_\mathbf{n}}{4\pi}$ with $N(0)$ the density of states (of one spin species) at the Fermi surface, $\epsilon^{\,}_c$ the cut-off energy and $\int\frac{d\Omega^{\,}_\mathbf{n}}{4\pi}$ the angular average over $\mathbf{n}$. The gap equation $1=\frac{g}{2}N(0)\int^{\epsilon^{\,}_c}_{-\epsilon^{\,}_c}\frac{d\xi^{\,}_\mathbf{k}}{E^{\,}_\mathbf{k}}$ has also been used. Combining Eqs.\,(\ref{Dk}), (\ref{Uk1}) and (\ref{Dk0}), we have $\mathbf{D}^{(1)}_{\mathbf{k},1}=\sum_{d=-1}^{3}\alpha^{\,}_d(R^d\mathbf{n})$ with the coefficient $\alpha^{\,}_d \in \mathbb{R}$. As mentioned, this again feeds back to $\braket{\mathbf{U}^{(1)}_\mathbf{k}}$ and therefore $\mathbf{D}^{(1)}_\mathbf{k}$. Finally, we have $\mathbf{D}^{(1)}_{\mathbf{k},\infty}=\sum_{d\in \mathbb{Z}}\alpha^{\,}_d(R^d\mathbf{n})$.
However, among all unit vectors $R^d\mathbf{n}$ ($d\in \mathbb{Z}$), maximally three of them are linearly independent, say $R\mathbf{n}$, $\mathbf{n}$, and $R^{-1}\mathbf{n}$. Therefore, $\mathbf{D}^{\,}_{\mathbf{k}}=\mathbf{D}^{(0)}_{\mathbf{k}}+\mathbf{D}^{(1)}_{\mathbf{k},\infty}$ in general takes the following form
\begin{align}\label{Dk2}
\mathbf{D}^{\,}_{\mathbf{k}}
=\frac{g^{\,}_D}{g}2\Delta\left[\frac{(\mathbf{n}\cdot R\mathbf{n})\mathbf{n}}{4}+aR\mathbf{n}+b\mathbf{n}+cR^{-1}\mathbf{n}\right],
\end{align}
with the coefficients $a$, $b$, and $c$ to be determined.

For computational convenience, we take the rotation axis of $R$ along the $\hat{z}$ direction and, as explained by Leggett \cite{LeggettEOM}, we will also set it as the direction of the magnetic field later. From Eqs.\,(\ref{Dk}), (\ref{Uk1}) and (\ref{Dk2}), after some lengthy but straightforward calculation we obtain again $\cos\theta=-1/4$, together with
\begin{align}
&a=-\frac{1}{56}+\left[\frac{c}{2}+\frac{3\sqrt{(\Delta/\epsilon^{\,}_c)^2+1}}{4gN(0)}\right],\\
&b=-\frac{1}{7}-3\left[\frac{c}{2}+\frac{3\sqrt{(\Delta/\epsilon^{\,}_c)^2+1}}{4gN(0)}\right].
\end{align}
It is worth mentioning that the precise meaning of $\theta=\theta^{\,}_L$ should be $\theta=\theta^{\,}_L+\mathcal{O}(g^{\,}_D/g)$. The freedom in choosing the first order term of $\theta$ leaves a free parameter among $a$, $b$, and $c$ \cite{abc}, consistent with that Eq.\,(\ref{HUk}) is only calculated up to $\mathcal{O}(g^2_D/g^2)$. We also mention in passing that there are two additional (higher energy) self-consistent solutions with $\cos\theta=\pm 1$, which we will not elaborate.


\begin{figure*}[t]
	\includegraphics[width=\textwidth]{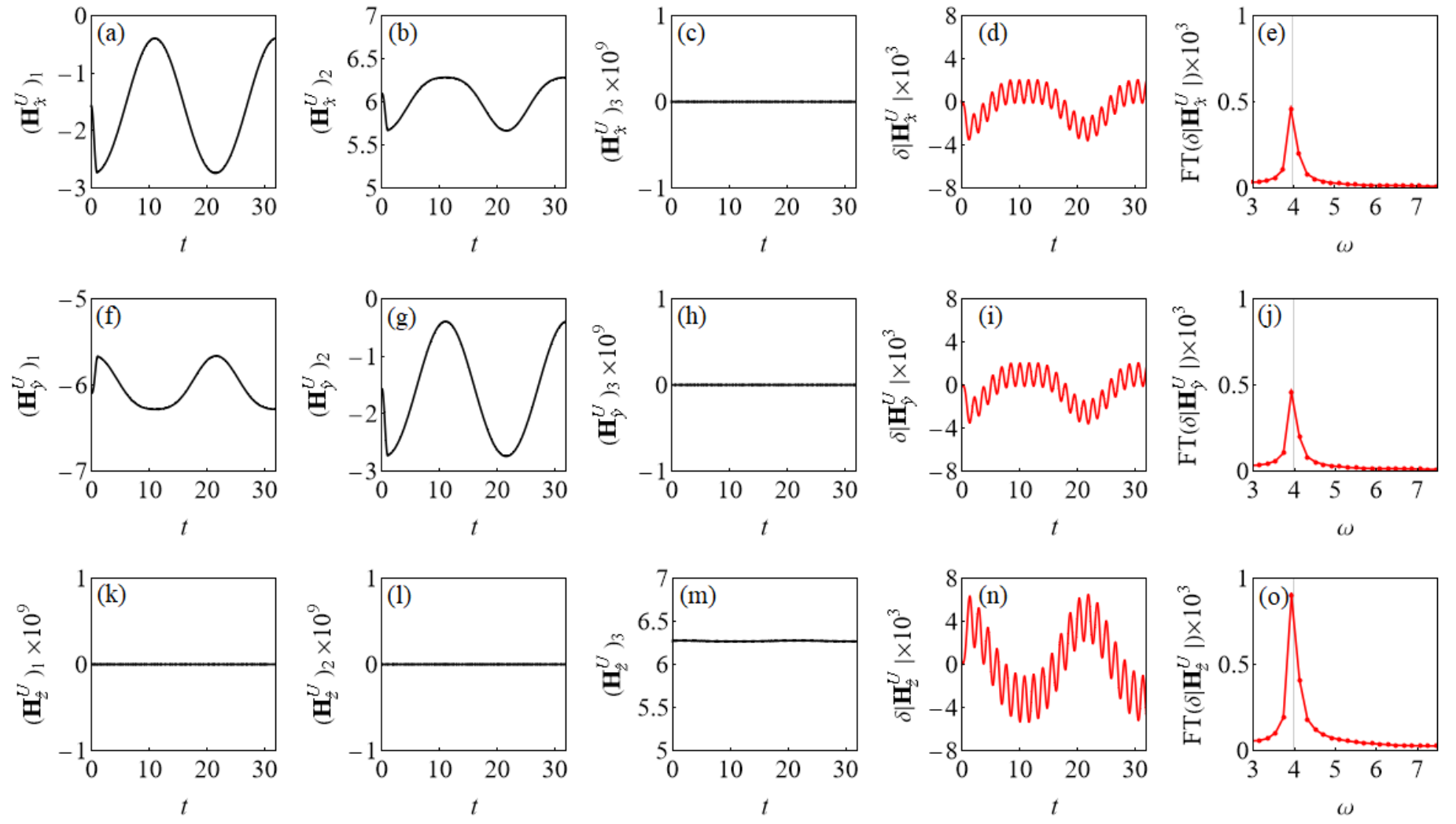}
	\caption{Quench dynamics of collective modes, with parameters $\Delta=\pi$, $g^{\,}_D/g=10^{-3}$, and $\mathbf{B}(t)=10^{-1}\pi\sin(\pi t) \Theta(t)\Theta(1-t)\hat{z}$, where $\Theta(t)$ is the unit step function. (a)-(c) are plots for the three components of $\mathbf{H}^U_{\hat{x}}$ after switching on the $\mathbf{B}(t)$ pulse. (d) is the amplitude deviation $\delta|\mathbf{H}^U_{\hat{x}}|(t)$ and (e) is its discrete Fourier transform [$\omega=2\pi(s-1)/32$ ($s\in\mathds{Z}$)], with a vertical grid line indicating the heavy Higgs boson at $\sqrt{8/5}\Delta$. (f)-(j) and (k)-(o) are similar plots for $\mathbf{H}^U_{\hat{y}}$ and $\mathbf{H}^U_{\hat{z}}$, as well as their amplitude deviations and Fourier transforms.}
	\label{Pump}
\end{figure*}

{\it Higgs boson pumping.---}Before turning on the magnetic field, it is instructive to have a dynamical perspective on the above process. Let us first note that $\mathbf{D}^{(1)}_{\mathbf{k}}$ in general is a function of time $t$ and it is appropriate to denote it as follows
\begin{align}
\mathbf{D}^{(1)}_{\mathbf{k}}(t)=\frac{g^{\,}_D}{g}2\Delta\left[\alpha(t)R+\beta(t)\mathds{1}^{\,}_3+\gamma(t)R^{-1}\right]\mathbf{n}.
\end{align}
Then, collective modes of the order parameter are associated with fluctuating components of $\alpha(t)R+\beta(t)\mathds{1}^{\,}_3+\gamma(t)R^{-1}\equiv F(t)$, which admits the tensor decomposition
\begin{align}
F(t)=R\left[f^{\,}_0(t)\mathds{1}^{\,}_3+f^{\,}_1(t)Q^{\,}_z+f^{\,}_2(t)Q^{\,}_{x^2+y^2-2z^2}\right],\\
Q^{\,}_z=\begin{bmatrix}
0 &-1 &0\\
1 &0 &0\\
0 &0 &0
\end{bmatrix},\quad
Q^{\,}_{x^2+y^2-2z^2}=\begin{bmatrix}
1 &0 &0\\
0 &1 &0\\
0 &0 &-2
\end{bmatrix},
\end{align}
with the coefficients $f^{\,}_0(t)=\alpha(t)+\frac{1}{6}\left[\beta(t)-\frac{3}{2}\gamma(t)\right]$, $f^{\,}_1(t)=-\frac{\sqrt{15}}{4}\left[\beta(t)-\frac{1}{2}\gamma(t)\right]$, and $f^{\,}_2(t)=-\frac{5}{12}[\beta(t)+\frac{3}{2}\gamma(t)]$. Now it becomes clear that three collective modes are possible, including the well-known light Higgs mode ($J=1$, $J^{\,}_z=0$) associated with $Q^{\,}_z$ and the $\sqrt{8/5}\Delta$ heavy Higgs mode ($J=2$, $J^{\,}_z=0$) associated with $Q^{\,}_{x^2+y^2-2z^2}$. The light Higgs mode corresponds to the oscillation of $\theta$ in the presence of the dipolar interaction. Indeed, we will see immediately that, under certain quench conditions, the light Higgs mode and the $\sqrt{8/5}\Delta$ heavy Higgs mode do exist and they are coupled. If we start from the self-consistent ground state that we determined for Eqs.\,(\ref{Uk}) and (\ref{HUk}), and turn on a suitable magnetic field, besides the expected light Higgs mode pumped by the magnetic field, the coupled $\sqrt{8/5}\Delta$ heavy Higgs mode should also be pumped. We next show that it is the case.

Let us start from Eq.\,(\ref{HMF}), from which we obtain ten coupled equations of motion for the SO(5) pseudo-spin. They have the same structure as Eq.\,(10) of Ref.\,\cite{Xu}, with the only difference coming from the definition of $\mathbf{H}^U_\mathbf{k}$ and $\mathbf{H}^V_\mathbf{k}$ in the presence of the dipolar interaction. In order to implement the dipolar interaction correctly, we expand Eq.\,(\ref{Rk}) in terms of Legendre polynomials
\begin{align}
&\mathbf{R}^U_\mathbf{k}=\sum_\mathbf{k'}V^D_{\mathbf{k}\mathbf{k'}}\left[\braket{\mathbf{U}^{\,}_\mathbf{k'}}-3\left(\mathbf{n'}\cdot\braket{\mathbf{U}^{\,}_\mathbf{k'}}\right)\mathbf{n'}\right],\\
&V^D_{\mathbf{k}\mathbf{k'}}=g^{\,}_D\left[(7/12)P^{\,}_3(\mathbf{n}\cdot\mathbf{n'})-(9/4)P^{\,}_1(\mathbf{n}\cdot\mathbf{n'})\right],
\end{align}
from which we see that the problem transforms to the integration of polynomials of degree six over the unit sphere. Thus, it is sufficient to handle the problem by employing the numerical method of 26-point Lebedev quadrature \cite{Lebedev} with the substitution
\begin{align}
\int\frac{d\Omega^{\,}_\mathbf{n}}{4\pi}f(\mathbf{n})=\sum_{j=1}^{3}\sum_{i=1}^{N^{\,}_j}A^{\,}_jf(a^j_i),
\end{align}
where we have coefficients $A^{\,}_1=1/21$, $A^{\,}_2=4/105$, and $A^{\,}_3=9/280$, together with $N^{\,}_1=6$, $N^{\,}_2=12$, and $N^{\,}_3=8$, totally $N=26$ points denoted as $a^1_i\in\{\pm\hat{x},\pm\hat{y},\pm\hat{z}\}$, $a^2_i\in\{\frac{\pm\hat{y}\pm\hat{z}}{\sqrt{2}},\frac{\pm\hat{x}\pm\hat{z}}{\sqrt{2}},\frac{\pm\hat{x}\pm\hat{y}}{\sqrt{2}}\}$, and $a^3_i\in\{\frac{\pm\hat{x}\pm\hat{y}\pm\hat{z}}{\sqrt{3}}\}$.

With the above recipe, we set $\xi^{\,}_\mathbf{k}=\xi/4$ with $\xi\in[-16,16]$ an integer. Then, each $\xi$ is associated with 26 Lebedev grid points $\{a^1_i,a^2_i,a^3_i\}$, and there is a total of $33\times26=858$ points sampled in momentum space. For a fixed $\mathbf{k}$, there are ten coupled equations of motion, and finally we need to solve 8580 coupled equations numerically. For visualization convenience, we choose $\Delta=\pi$ and $g^{\,}_D/g=10^{-3}$. As a result, after turning on a magnetic pulse $\mathbf{B}(t)=10^{-1}\pi\sin(\pi t) \Theta(t)\Theta(1-t)\hat{z}$, with $\Theta(t)$ the unit step function, we have the real-time quench dynamics of all nine components of $\mathbf{H}^U_{\hat{x}}$, $\mathbf{H}^U_{\hat{y}}$, and $\mathbf{H}^U_{\hat{z}}$ plotted in Fig.\,\ref{Pump}\,(a)-(c), (f)-(h), and (k)-(m), which characterize collective modes on top of the ground state. As one can see, there are five nonvanishing components, i.e., Fig.\,\ref{Pump}\,(a), (b), (f), (g), and (m), with the first four mainly associated with the light Higgs mode if one views them as $\cos[\theta(t)]$, $\sin[\theta(t)]$, $-\sin[\theta(t)]$, and $\cos[\theta(t)]$, respectively. To identify the heavy Higgs mode, we plot in Fig.\,\ref{Pump}\,(d) the amplitude deviation $\delta|\mathbf{H}^U_{\hat{x}}|(t)$ and in Fig.\,\ref{Pump}\,(e) its discrete Fourier transform $\frac{1}{320}\sum_{\ell=1}^{320}\delta|\mathbf{H}^U_{\hat{x}}|(t^{\,}_{\ell-1})e^{\i\omega t^{\,}_{\ell-1}}$, where $t^{\,}_{\ell-1}=0.1(\ell-1)$ and $\omega=2\pi(s-1)/32$ with $s$ integers. Evidently, modulated by the low-frequency light Higgs background, the high-frequency oscillation has a frequency peak located around $\sqrt{8/5}\Delta$. The same scenario happens to $\delta|\mathbf{H}^U_{\hat{y}}|(t)$ [see Fig.\,\ref{Pump}\,(i) and (j)] and $\delta|\mathbf{H}^U_{\hat{z}}|(t)$ [see Fig.\,\ref{Pump}\,(n) and (o)]. It is not hard to see that $\delta|\mathbf{H}^U_{\hat{z}}|(t)$ is twice the amplitude of $\delta|\mathbf{H}^U_{\hat{x}}|(t)$ and $\delta|\mathbf{H}^U_{\hat{y}}|(t)$, but oscillates out of phase with them, all consistent with the characteristics of the irreducible representation $Q^{\,}_{x^2+y^2-2z^2}$. Last but not the least, since the first two components of $\mathbf{H}^U_{\hat{z}}$ vanish, the light Higgs background in the dynamics of $\delta|\mathbf{H}^U_{\hat{z}}|(t)$ is certainly associated with fluctuations of $(\mathbf{H}^U_{\hat{z}})^{\,}_3$, which originally has nothing to do with spin-wave modes if the dipolar interaction is absent. Thus, we reach the conclusion that the $\sqrt{8/5}\Delta$ heavy Higgs mode with $J=2$ and $J^{\,}_z=0$, accompanying the well-known light Higgs mode, has been excited by the magnetic pulse with no doubt.

{\it Concluding remarks.---}We have investigated the interplay between a $\sqrt{8/5}\Delta$ heavy Higgs boson and a light Higgs boson on top of the BW ground state, employing an effective SO(5) pseudo-spin model including the dipolar interaction. As a prerequisite, the ground state is solved self-consistently to the first order in the dipolar interaction strength, through the mathematical method of iteration. After turning on a suitable magnetic pulse, along with the light Higgs boson ($J=1$, $J^{\,}_z=0$), the $\sqrt{8/5}\Delta$ heavy Higgs boson ($J=2$, $J^{\,}_z=0$) is successfully excited and the two are coupled. Just like the zero sound absorption at $\sqrt{8/5}\Delta$ \cite{Lee,Halperin}, the coupled oscillation of the heavy and light Higgs bosons is also a small effect but may still be observable after resonant amplification. What's more, it opens up the possibility to observe the light Higgs boson pair production \cite{Volovik2016} at the high energy scale, reflecting the parametric decay of the heavy Higgs boson (to the light ones with large momenta) and, hopefully, shedding light on high energy physics \cite{Volovik2014,Higgsbosonpair}.

{\it Acknowledgments.---}The author would like to thank C. Wu for comments on this work. The author also thanks G. Ortiz and G. E. Volovik for discussions on an early work of Higgs bosons.


\begin{thebibliography}{}
	
\bibitem{Leggett}
A. J. Leggett,
{A theoretical description of the new phases of liquid $^3$He},
Rev. Mod. Phys. {\bf 47}, 331 (1975).

\bibitem{BW}
R. Balian and N. R. Werthamer,
{Superconductivity with pairs in a relative $p$ wave},
Phys. Rev. {\bf 131}, 1553 (1963).

\bibitem{LeggettSBSOS}
A. J. Leggett,
{Interpretation of Recent Results on He$^3$ below 3 mK: A New Liquid Phase?},
Phys. Rev. Lett. {\bf 29}, 1227 (1972).

\bibitem{Sauls}
J. A. Sauls,
{On the excitations of a Balian--Werthamer superconductor},
J. Low Temp. Phys. {\bf 208}, 87 (2022).

\bibitem{J}
The twisted angular momentum operator (see Ref.\,[\onlinecite{Hughes}] for example) can be written as $\mathbf{J}=\mathbf{L}+R^{-1}\mathbf{S}$, with $\mathbf{L}$ the orbital angular momentum, $\mathbf{S}$ the spin angular momentum, and $R$ the rotation matrix defining the BW state.

\bibitem{Hughes}
T. I. Tuegel and T. L. Hughes,
{Hall viscosity and the acoustic Faraday effect},
Phys. Rev. B {\bf 96}, 174524 (2017).

\bibitem{LeggettEOM}
A. J. Leggett,
{Microscopic Theory of NMR in an Anisotropic Superfluid ($^3$He \textit{A})},
Phys. Rev. Lett. {\bf 31}, 352 (1973).

\bibitem{Volovik2014}
G. E. Volovik and M. A. Zubkov,
{Higgs bosons in particle physics and in condensed matter},
J. Low Temp. Phys. {\bf 175}, 486 (2014).

\bibitem{Volovik2016}
V. V. Zavjalov, S. Autti, V. B. Eltsov, P. J. Heikkinen, and G. E. Volovik,
{Light Higgs channel of the resonant decay of magnon condensate in superfluid $^3$He-B},
Nat. Commun. {\bf 7}, 10294 (2016).

\bibitem{Wolfle}
P. W\"olfle,
{Collisionless collective modes in superfluid $^3$He},
Physica B {\bf 90}, 96 (1977).

\bibitem{Xu}
Q.-R. Xu and G. Ortiz,
{Quantum quenches of an SO(5) pseudospin reveal Higgs bosons},
Phys. Pev. B {\bf 107}, L100503 (2023).

\bibitem{Volkov}
A. F. Volkov and Sh. M. Kogan,
{Collisionless relaxation of the energy gap in superconductors},
Sov. Phys. JETP {\bf 38}, 1018 (1974).

\bibitem{Yuzbashyan}
E. A. Yuzbashyan, O. Tsyplyatyev, and B. L. Altshuler,
{Relaxation and Persistent Oscillations of the Order Parameter in Fermionic Condensates},
Phys. Rev. Lett. {\bf 96}, 097005 (2006).

\bibitem{Matsunaga}
R. Matsunaga, Y. I. Hamada, K. Makise, Y. Uzawa, H. Terai, Z. Wang, and R. Shimano,
{Higgs Amplitude Mode in the BCS Superconductors Nb$^{\,}_{1-x}$Ti$^{\,}_x$N Induced by Terahertz Pulse Excitation},
Phys. Rev. Lett. {\bf 111}, 057002 (2013).

\bibitem{Tsuji}
N. Tsuji and H. Aoki,
{Theory of Anderson pseudospin resonance with Higgs mode in superconductors},
Phys. Rev. B {\bf 92}, 064508 (2015).

\bibitem{Shimano}
R. Shimano and N. Tsuji,
{Higgs mode in superconductors},
Annu. Rev. Condens. Matter Phys. {\bf 11}, 103 (2020).

\bibitem{Hasegawa}
Y. Hasegawa, T. Usagawa, and F. Iwamoto,
{Application of the 5-dimensional spin to the theory of superfluid $^3$He},
Prog. Theor. Phys. {\bf 62}, 1458 (1979).

\bibitem{abc}
It has been tested that the free parameter has negaligible effects on the final result of Fig.\,\ref{Pump} as long as $\mathbf{D}^{\,}_\mathbf{k}$ is a perturbation to $\mathbf{H}^U_\mathbf{k}$.

\bibitem{Lebedev}
V. I. Lebedev,
{Values of the nodes and weights of ninth to seventeenth order Gauss-Markov quadrature formulae invariant under the octahedron group with inversion},
USSR Comput. Math. Math. Phys. {\bf 15}, 44 (1975).

\bibitem{Lee}
R. W. Giannetta, A. Ahonen, E. Polturak, J. Saunders, E. K. Zeise, R. C. Richardson, and D. M. Lee,
{Observation of a New Sound-Attenuation Peak in Superfluid $^3$He-B},
Phys. Rev. Lett. {\bf 45}, 262 (1980).

\bibitem{Halperin}
D. B. Mast, B. K. Sarma, J. R. Owers-Bradley, I. D. Calder, J. B. Ketterson, and W. P. Halperin,
{Measurements of High-Frequency Sound Propagation in $^3$He-B},
Phys. Rev. Lett. {\bf 45}, 266 (1980).

\bibitem{Higgsbosonpair}
M. Gouzevitch and A. Carvalho,
{A review of Higgs boson pair production},
Rev. Phys. {\bf 5}, 100039 (2020).
	
\end{thebibliography}
\end{document}